\documentclass{elsart}

\usepackage{amsmath,amsfonts,amssymb}
\usepackage{mathrsfs,bbm}                               
\usepackage{graphicx,subfigure,pst-all,psfrag}		

\usepackage[sort&compress]{natbib}			
\bibpunct{[}{]}{,}{n}{}{;}				

	\newcommand{\E}[1]{\ensuremath{\mathbb{E}\!\left[#1\right]}}

	\newcommand{\abs}[1]{\ensuremath{\left|#1\right|}}

	\renewcommand{\Re}{\ensuremath{\mathbb{R}}}

	\renewcommand{\thefootnote}{\fnsymbol{footnote}}
	\DeclareMathAlphabet{\mathpzc}{OT1}{pzc}{m}{it}
\begin{document}
\begin{frontmatter}
\title{Characterization of laser propagation through turbulent media by quantifiers based on the wavelet transform: dynamic study} 
\author[ciop]{L. Zunino\corauthref{cor}},
\corauth[cor]{Corresponding author.}
\ead{lucianoz@ciop.unlp.edu.ar}
\author[pucv]{D. G. P\'erez},
\ead{dario.perez@ucv.cl}
\author[ciop,fis]{M. Garavaglia}
\ead{garavagliam@ciop.unlp.edu.ar}
and
\author[uba]{Osvaldo A.  Rosso}
\ead{oarosso@fibertel.com.ar}

\address[ciop]{Centro de Investigaciones \'Opticas (CIOp), CC. 124 Correo Central,1900 La Plata, Argentina.}

\address[pucv]{Instituto de F\'isica, Pontificia Universidad Cat\'olica de Valpara\'iso (PUCV), 23-40025 Valpara\'iso, Chile.}

\address[fis]{Departamento de F\'{\i}sica, Facultad de Ciencias Exactas, Universidad Nacional de La Plata (UNLP), 1900 La Plata, Argentina.}
                 
\address[uba]{Instituto de C\'alculo, Facultad de Ciencias Exactas y Naturales, Universidad de Buenos Aires (UBA), Pabell\'on II, Ciudad Universitaria, 1428 Ciudad de Buenos Aires, Argentina. }


\begin{abstract} 
We analyze, within the wavelet theory framework, the wandering over a screen of the centroid of a laser beam after it has propagated through a time-changing laboratory-generated turbulence. Following a previous work (Fractals 12 (2004) 223) two quantifiers are used, the Hurst parameter, $H$, and the Normalized Total Wavelet Entropy, $\text{NTWS}$. The temporal evolution of both quantifiers, obtained from the laser spot data stream is studied and compared. This allows us to extract information of the stochastic process associated to the turbulence dynamics.
\end{abstract}

\end{frontmatter}

\section{INTRODUCTION}
\label{sec:intro}

The purpose of this work is to statistically describe laser beam propagation through time-changing laboratory-generated turbulence. To do so, we analyze data stream corresponding to the centroid position of the laser spot by using two different quantifiers obtained from the wavelet theory: the \textit{Hurst parameter}, $H$, and the \textit{Normalized Total Wavelet Entropy}, $\text{NTWS}$. The former quantifier results from modeling the centroid's coordinates as a \textit{fractional Brownian motion} (fBm) at stationary turbulence strength~\cite{paper:zunino2004}, while the latter has been used for a wider set of stochastic processes---see Ref.~\cite{paper:rosso2001}.

The fBm was discovered by Kolmogorov~\cite{paper:kolmogorov1940} and defined by Mandelbrot and Van Ness~\cite{paper:mandelbrot1968} as the only one family of processes which are gaussian, self-similar, and with stationary increments. The normalized family of these gaussian processes, $B^H$, is the one with $B^H(0)=0$ almost surely, $\mathbb{E}[B^H(t)]=0$, and covariance
\begin{equation}
\E{B^H(t)B^H(s)}=
\frac{1}{2}\left(\abs{t}^{2H}+\abs{s}^{2H}-\abs{t-s}^{2H}\right),
\end{equation}
for $s,t\in\Re$. The power exponent $H$ is also known as scaling exponent and its range is bounded between $0$ and $1$. The estimation of this parameter plays a key role modeling a fBm time series. One remarkable property of this family $B^H$ is that the $H$ parameter regulates the presence or absence of memory. In fact, it can be separated in three subfamilies accordingly: long-memory for $1/2<H<1$, memoryless at $H=1/2$ (ordinary Brownian motion), and short-memory in the case $0<H<1/2$. Likewise, the Hurst parameter can be thought as the probability of the next increment of the signal have the same sign as the previous increment. Thus, it tunes the trajectory regularity. Fractional Brownian motions are continuous but non-differentiable processes (in the usual sense), and only give generalized spectra $1/f^\alpha$ with exponents $\alpha$ between $1$ and $3$. As a nonstationary process, the fBm does not have a spectrum defined in the usual sense; however, it is possible to  define a power spectrum of the form~\cite{paper:perez2001}:
\begin{equation}
S_{B^H}(f)={\frac{1}{\left|f\right|^{2H+1}}}. 
\label{eq-4}
\end{equation}
Remember that this equation is not a valid power spectrum in the theory of stationary processes since it is a non-integrable function in the classical sense.

Several properties evidence that wavelet analysis is well-suited to fBm:
\begin{enumerate}
\item fBm is nonstationary but the wavelet coefficients form a stationary process at 
each scale~\cite{paper:flandrin1989,paper:flandrin1992};
\item fBm exhibits positive long-range correlation in the range $1/2<H<1$ but wavelet coefficients have a correlation which is highly small as soon as $N>H+1/2$, where $N$ is the number of vanishing moments associated to the mother wavelet $\psi(t)$~\cite{paper:flandrin1992,paper:tewfik1992};
\item the self-similarity of fBm is reproduced in its wavelet coefficients, 
whose variance varies as a  power law as a function of scale $j$~\cite{paper:flandrin1989,paper:flandrin1992}
\begin{equation}
\log_2 \left\{\mathbb{E}\left[C^2_j(k)\mid_{B^H}\right]\right\} \propto-(2H+1)j. 
\label{eq2:1}
\end{equation}
\end{enumerate}
These features are due to the fact that the wavelet family $\psi_{a,b}$ is generated by dilations and translations of a unique admissible mother wavelet $\psi(t)$. So, the family itself exhibits scale invariance. 
It should be noted that the first two properties are valid for any process with stationary increments~\cite{paper:masry1993}.

In particular these properties are widely used for estimating $H$ or the related spectral exponent $\alpha=2H+1$~\cite{paper:abry1998,paper:perez2001,paper:soltani2004}. Through the \textit{Logscale Diagram} the threefold objective: detection, identification and measurements of the scaling exponent can be achieved~\cite{paper:abry2000}. Basically the estimation problem turns into a linear regression slope estimation.

In order to model the time-changing turbulence situation we consider a generalization where the parameter $H$ is no longer constant, but a continuous function of the time $t$ ($H \to H(t)$). This generalization was introduced in financial research to model the behaviour of stock market index time series~\cite{paper:grech2004,paper:cajueiro2004,paper:carbone2004}. Also, it was recently used to characterize dynamic speckle or biospeckle~\cite{paper:passoni2004}. A single scaling exponent would be unable to show the complex dynamics inherent to the data. The constraint of stationary increments is relaxed in this case. \textit{Multifractional Brownian motion} (mBm)~\cite{paper:peltier1995} was formalized as a class of processes which satisfies these properties.

\begin{figure}[!h]
\begin{center}
\psfrag{signal}[][]{signal}
\psfrag{alpha}[][t]{$\alpha$}
\psfrag{ntws}[][t]{NTWS}
\psfrag{window}[][b]{window number}
\psfrag{time}[][b]{data number}
     \hspace{-.2cm}\subfigure{
          \includegraphics[height=.358\textheight]{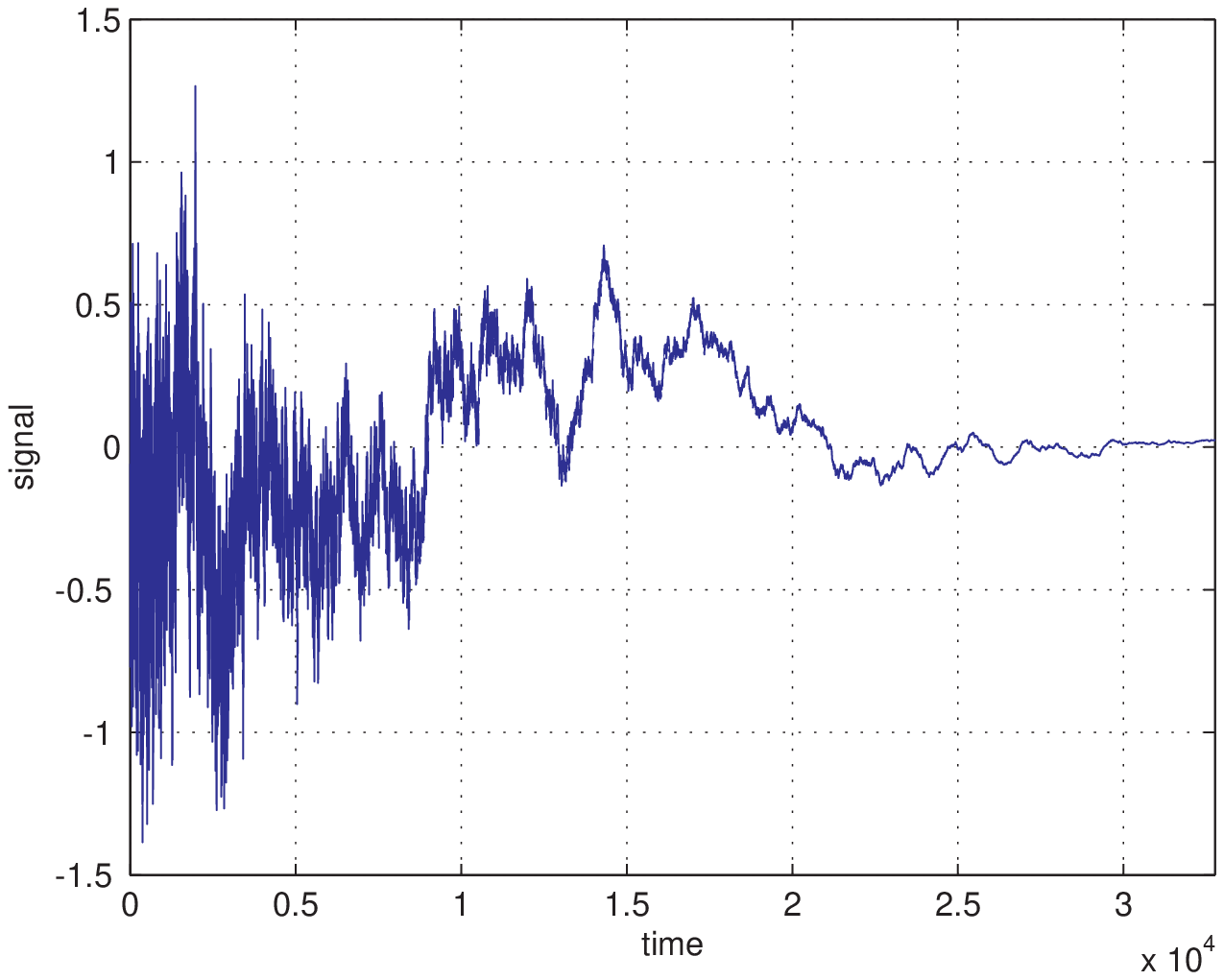}}
          
     \hspace{.1cm}\subfigure{
          \includegraphics[height=.372\textheight]{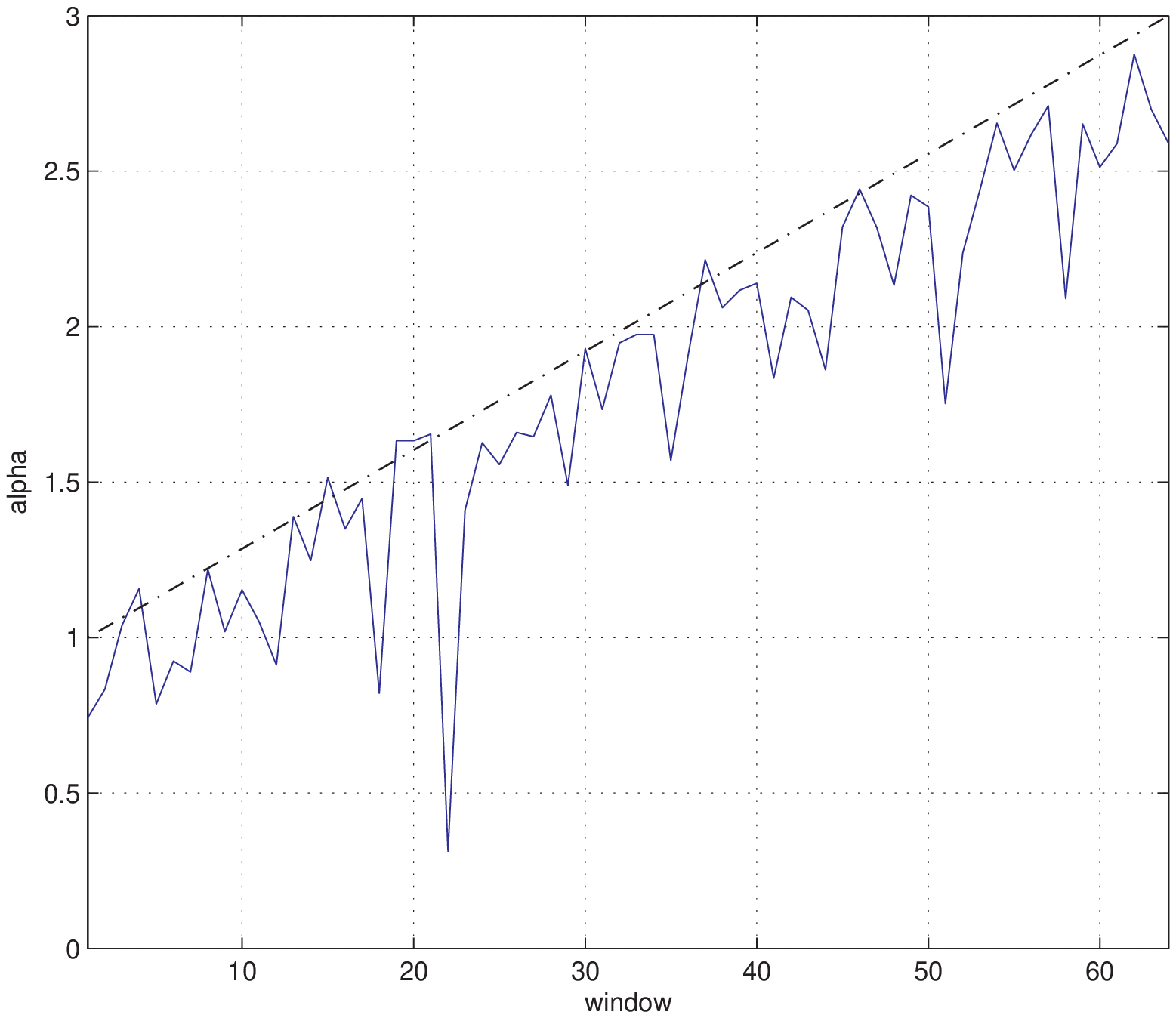}}

     \subfigure{
          \includegraphics[height=.35\textheight]{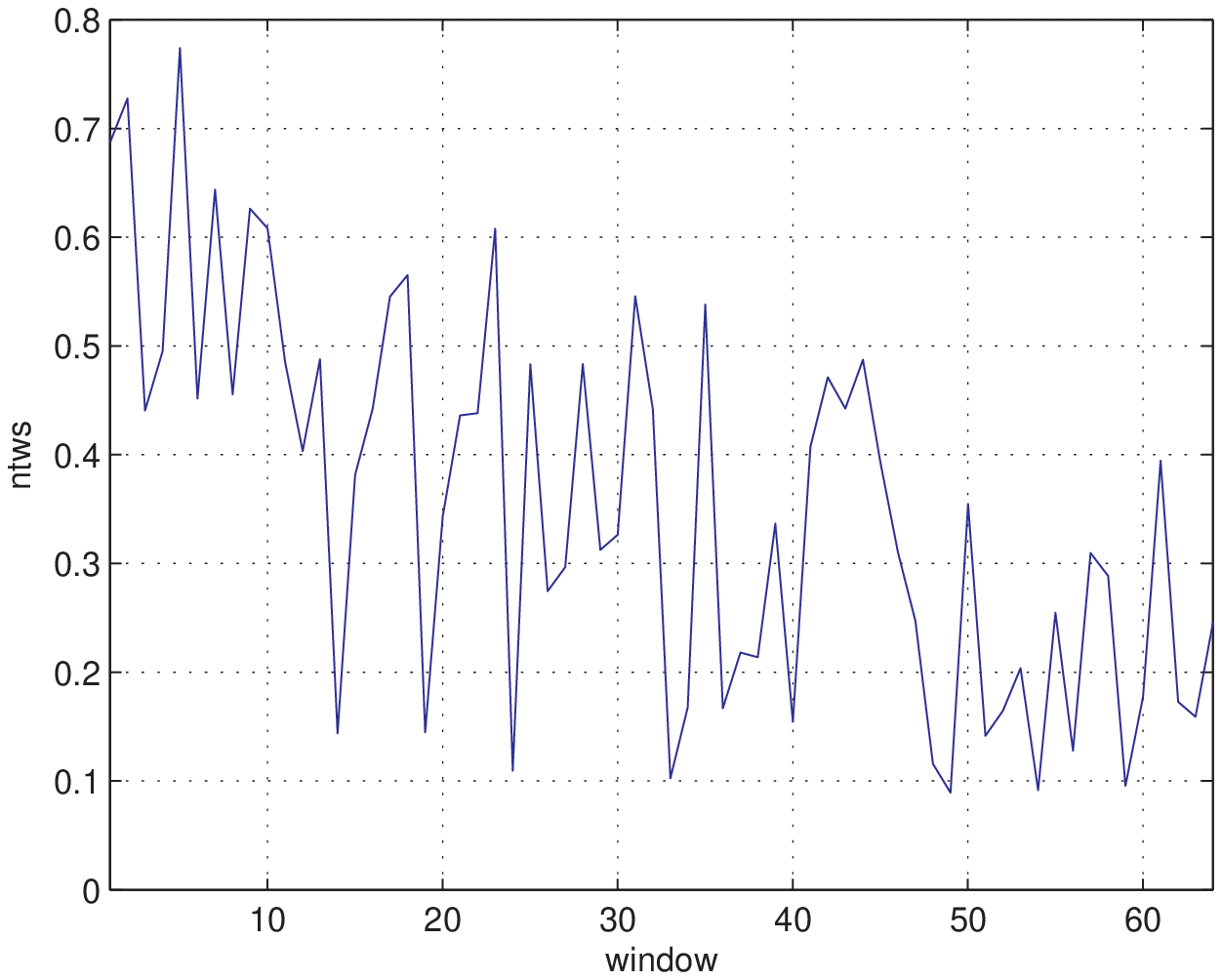}}
\caption{ Top: Synthetically generated mBm signal with $H$ changing linearly from $0.1$ to $0.9$  with $t$. Middle: Theoretical (dashed curve) and measured (continuous curve) Hurst parameter for this simulated mBm. Bottom: Measured $\text{NTWS}$.\label{figure:1}}
\end{center}
\end{figure}

We calculate the time-dependent Hurst exponent by using the wavelet properties. Provided that variations of $H$ are smooth enough, the signal is divided into $i$ non-overlapping temporal windows and the scaling exponent is calculated for each subset according to the procedure described at Ref.~\cite{paper:zunino2004}. A sequence of Hurst parameter values is obtained. They give the local scaling exponent around a given instant of time. Artificially mBm were analyzed in order to test the quality of our estimator.  In Figure \ref{figure:1} one can compare the theoretical and experimental results for a mBm with $H$ changing linearly from $0.1$ to $0.9$  with $t$. The Matlab code introduced by J. F. Coeurjolly~\cite{phd:coeurjolly2000} was implemented to simulate the mBm. The signal was divided in 64 temporal windows of 512 data points. We used the orthogonal cubic spline functions as mother wavelet and the resolution levels from $j = -9$ to $j = -1$.

At the same time, the $\text{NTWS}$ is also applied to study this time-changing turbulence. Introduced as a measure of the degree of order-disorder of the signal~\cite{paper:rosso2001}, it provides information about the underlying dynamical process associated with the signal. We define the $\text{NTWS}$ as
\begin{equation}
\label{eq:wav9}
S_\text{WT}=  -\sum_{j=-N}^{-1}  p_j  \cdot \ln p_j/ S^ \text{max},
\end{equation}
where
\begin{equation}
\label{eq:wav10}
S^ \text{max}=\ln N
\end{equation}
with $N$, at least, the base 2 logarithm of the number of data points and $\{p_j\}$ represent the \textit{Relative Wavelet Energy} (RWE). These are defined as
\begin{equation}
\label{eq:wav8}
p_j=\mathcal{E}_j/\mathcal{E}_{ \textit{tot}},
\end{equation}
with $\mathcal{E}_j=\mathbb{E}[C^2_j(k)]$ the energy at each resolution levels $j = -N,\cdots ,-2, -1$ and $\mathcal{E}_{ \text{tot}}=\sum_{j<0}\mathcal{E}_j$. They yield, at different scales, the probability distribution for the energy. It should be remarked that an orthogonal mother wavelet must be used within this theory---further details can be found at Ref.~\cite{paper:zunino2004}. Indeed, a very ordered process can be represented by a signal with a narrow band spectrum. A wavelet representation of such a signal will be resolved in few wavelet resolution levels, i. e., all RWE will be (almost) zero except at the wavelet resolution levels which includes the representative signal frequency. For this special levels the RWE will be almost equal to one. As a consequence, the NTWS will acquire a very small, vanishing value. A signal generated by a totally random process or chaotic one can be taken as representative of a very disordered behavior. This kind of signal will have a wavelet representation with significant contributions coming from all frequency bands. Moreover, one could expect that all contributions will be of the same order. Consequently, the RWE will be almost equal at all resolutions levels, and the NTWS will acquire its maximum possible value. Higher values for wavelet entropy means higher dynamical complexity, higher irregular behaviour and, of course, lower predictability.

The time evolution of $\text{NTWS}$ can be easily implemented. So, it is widely used to study a wide set of nonstationary natural signal. In particular, it was introduced to quantify the degree of disorder in the electroencephalographic epileptic records giving information about the underlying dynamical process in the brain~\cite{paper:rosso2002}, more specifically of the synchrony of the group cells involved in the different neural responses. Also, monthly time series of different solar activity indices (sunspot numbers, sunspot areas and flare index) were analyzed~\cite{paper:sello2000,paper:sello2003}. The disorder content of solar cycle activity can be derived by analyzing the wavelet entropy time evolution. Likewise, the dynamic speckle phenomenon mentioned above has also been analyzed by using these wavelet based entropy concepts~\cite{paper:passoni2004a}. In a recent paper it is investigated the relation existing between these two quantifiers---$H$ and $\text{NTWS}$---when they are used for analyzing fBm \cite{paper:perez2005}. Figure \ref{figure:1} (top and bottom) shows the mBm and his corresponding NTWS, where the same temporal windows, mother wavelet and resolution levels were used.

\section{EXPERIMENTAL SETUP AND DATA ADQUISITION}
\label{sec:exp}

The experimental measures were performed in a laboratory by producing thermal convective turbulence with two electrical heaters in a row. Three different turbulence intensities were generated changing the amount of heat dissipated for each electrical heater: normal, soft and hard turbulence. Along the laser path three electronic thermometers sense the air temperature---$T1, T2$ and $T3$, see Fig. \ref{figure:2} (bottom). A time  series corresponding to the fluctuations of the centroid position of a laser beam's spot (wandering) over a screen, after propagation through this time-changing laboratory-generated turbulence, were recorded  with a  position sensitive detector located as screen at the end of the path. This record consists of $2,500,000$ spot beam centroid coordinates measurements with $500,000$ data for each laboratory-generated turbulence condition. Further details of the experiment can be found at Ref.~\cite{paper:zunino2004}. The temperature and signal records can be observed in Figure \ref{figure:2}. There, it can be observed that the turbulence is increased, and subsequently decreased up to recover the initial situation. 

\section{RESULTS AND CONCLUSIONS}
\label{sec:res}

In the present work, we employ orthogonal cubic spline functions as mother
wavelets. Among several alternatives, cubic spline functions are symmetric and combine
in a suitable proportion smoothness with numerical advantages. They have become a recommendable tool for representing natural signals~\cite{paper:unser1999,paper:thevenaz2000}.
The signal was divided into $606$ non-overlapping temporal windows of 4096 data points. Resolution levels  between $j=-7$ and $j=-3$ were used to calculate both quantifiers. The first two levels ($j=-1$ and $j=-2$) were dropped to reduce the noise introduced by the sistem, while the lower levels were excluded to reduce nonstationary effects as commented in Ref.~\cite{paper:zunino2004}.

\begin{figure}[!h]
\begin{center}
\psfrag{signal}[][t]{signal}
\psfrag{coordx}[][t]{coordinate $x$}
\psfrag{coordy}[][t]{coordinate $y$}
\psfrag{temprecord}[][t]{temperature record}
\psfrag{archive}[][b]{archive number}
\psfrag{time}[][b]{time}
\psfrag{temp}[][t]{Temperature $[0.1{}^\circ\text{C}]$}
     \subfigure{
          \includegraphics[height=.35\textheight]{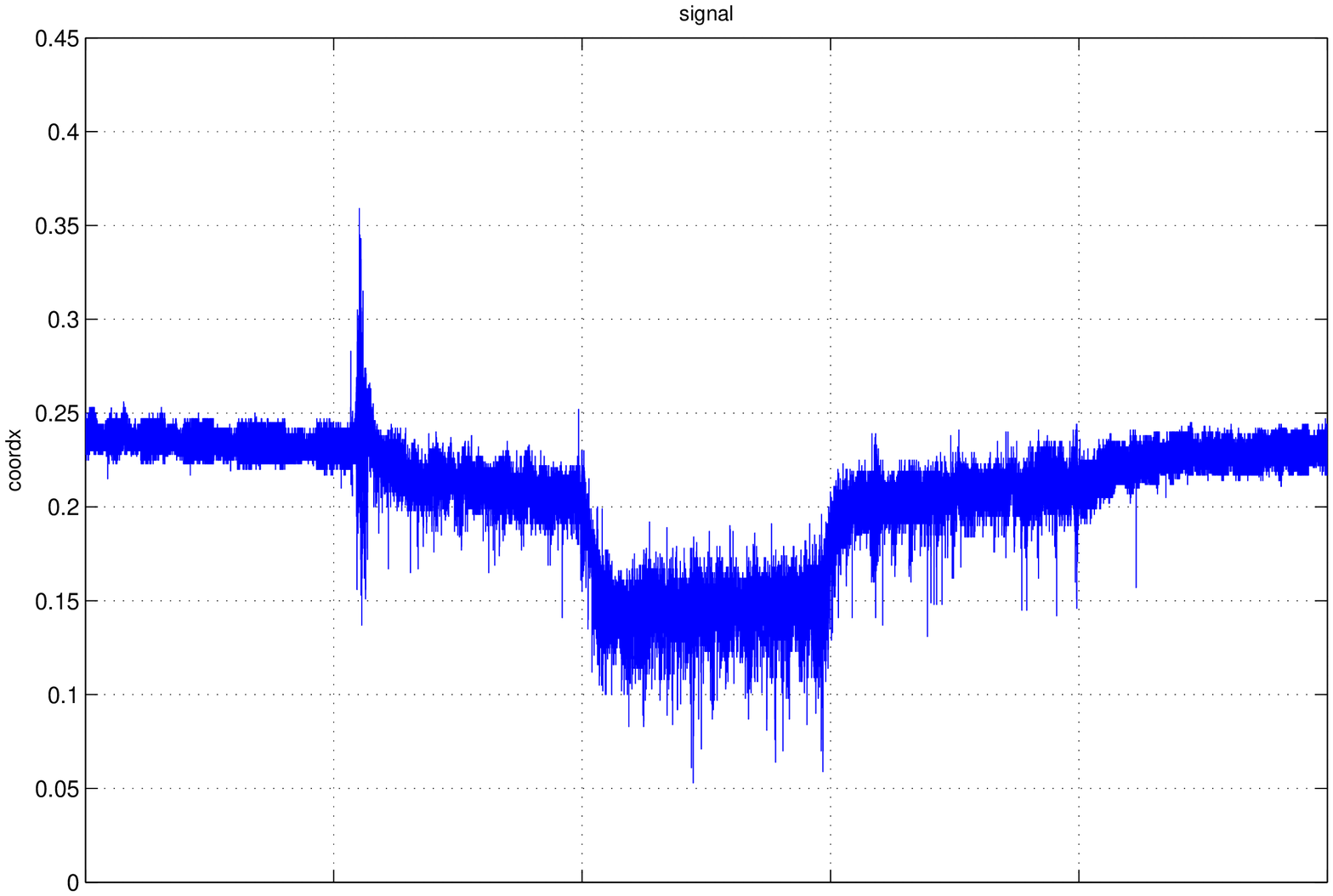}}
          
     \vspace{-1cm}
     \subfigure{
          \includegraphics[height=.35\textheight]{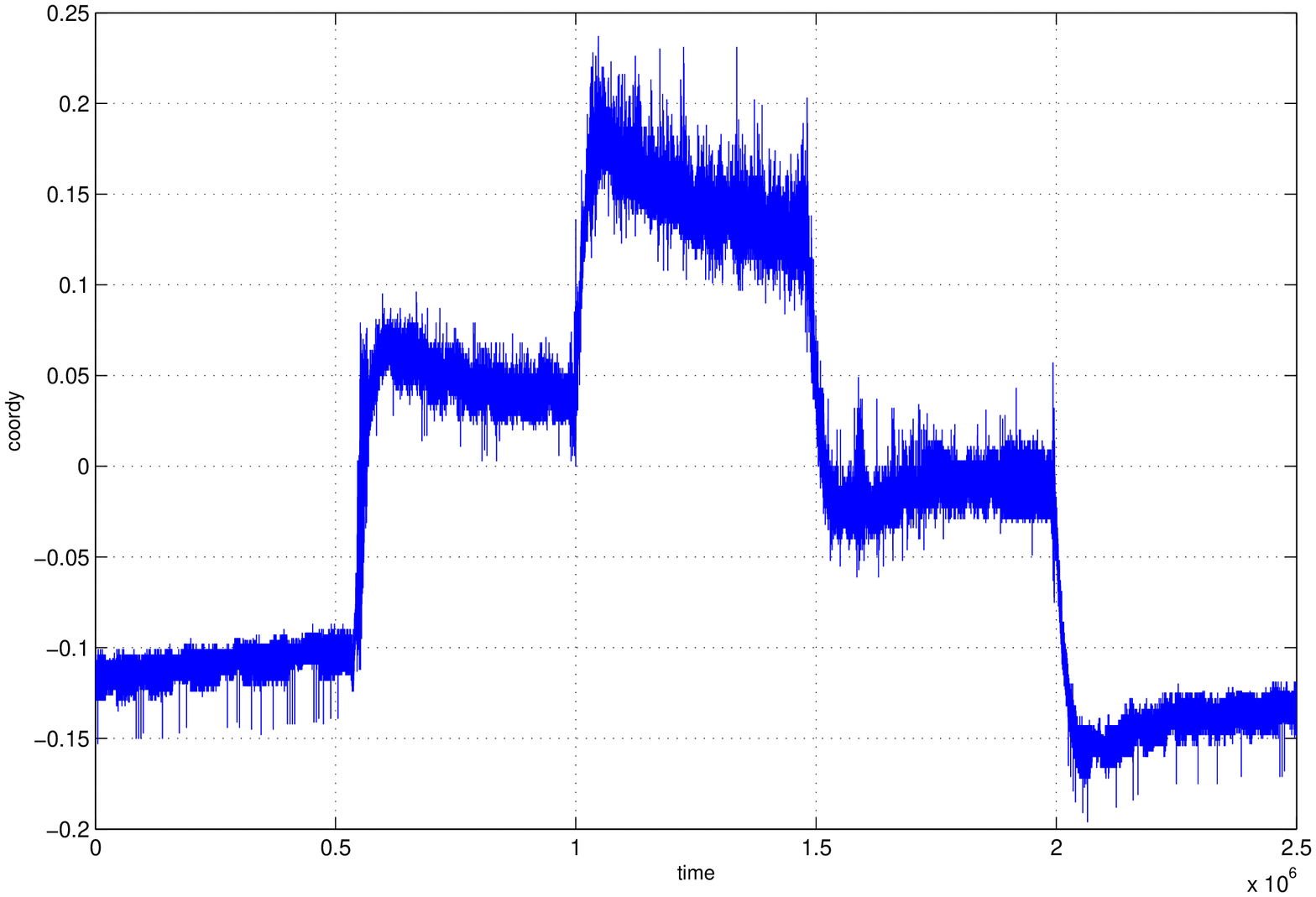}}

     \subfigure{
          \includegraphics[height=.35\textheight]{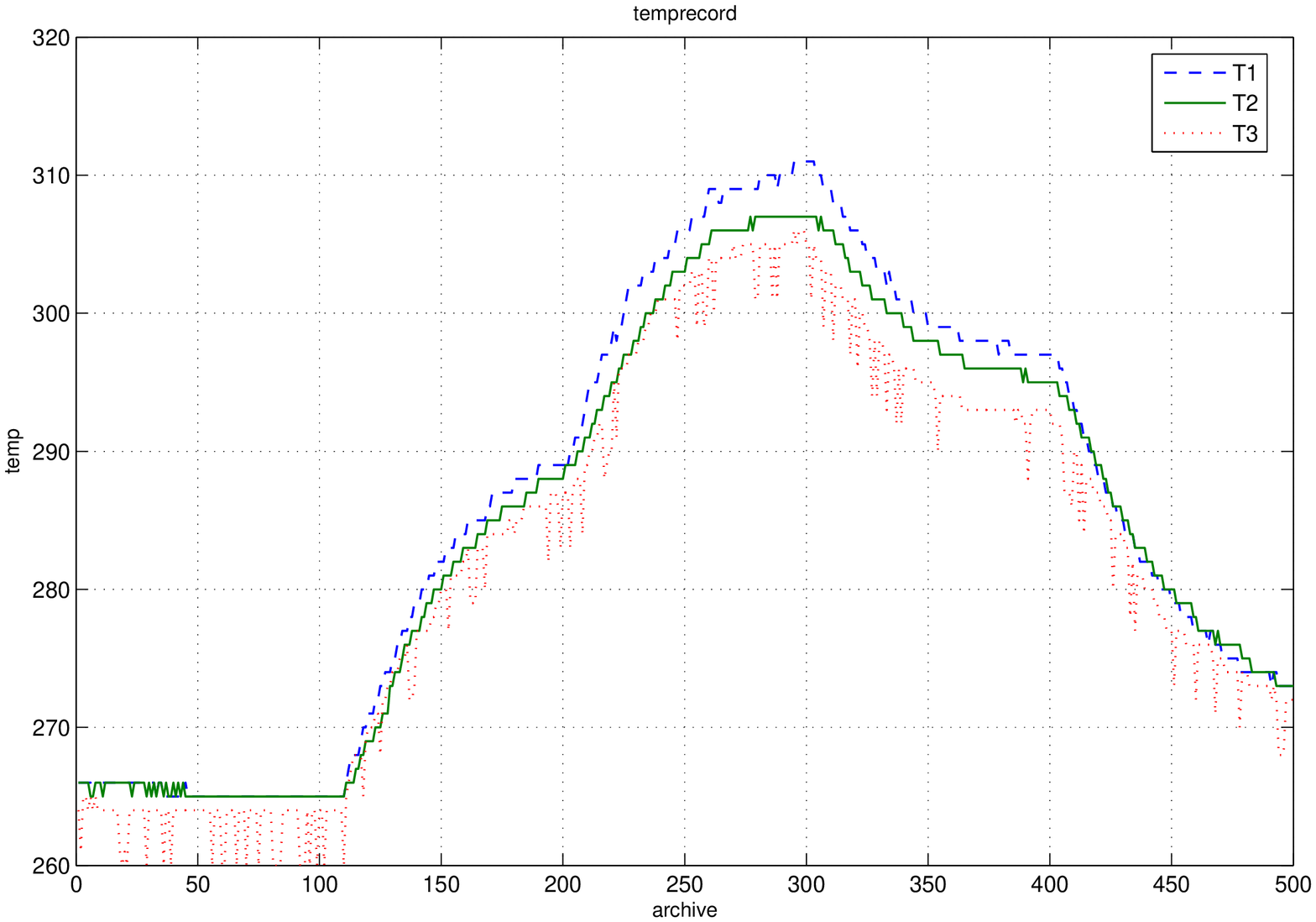}}
\caption{ Experimental records for the $x$ (top) and $y$ (middle) coordinates and the associated temperature record (bottom). \label{figure:2}}
\end{center}
\end{figure}

\begin{figure}[!h]
\psfrag{Coordinate x}[][t]{coordinate $x$}
\psfrag{coordy}[][t]{coordinate $y$}
\psfrag*{window}[][b]{window number}
\psfrag{alpha}[][t]{$\alpha$}
\psfrag{NTWS}[][t]{NTWS}
     \subfigure{
          \includegraphics[height=.25\textheight]{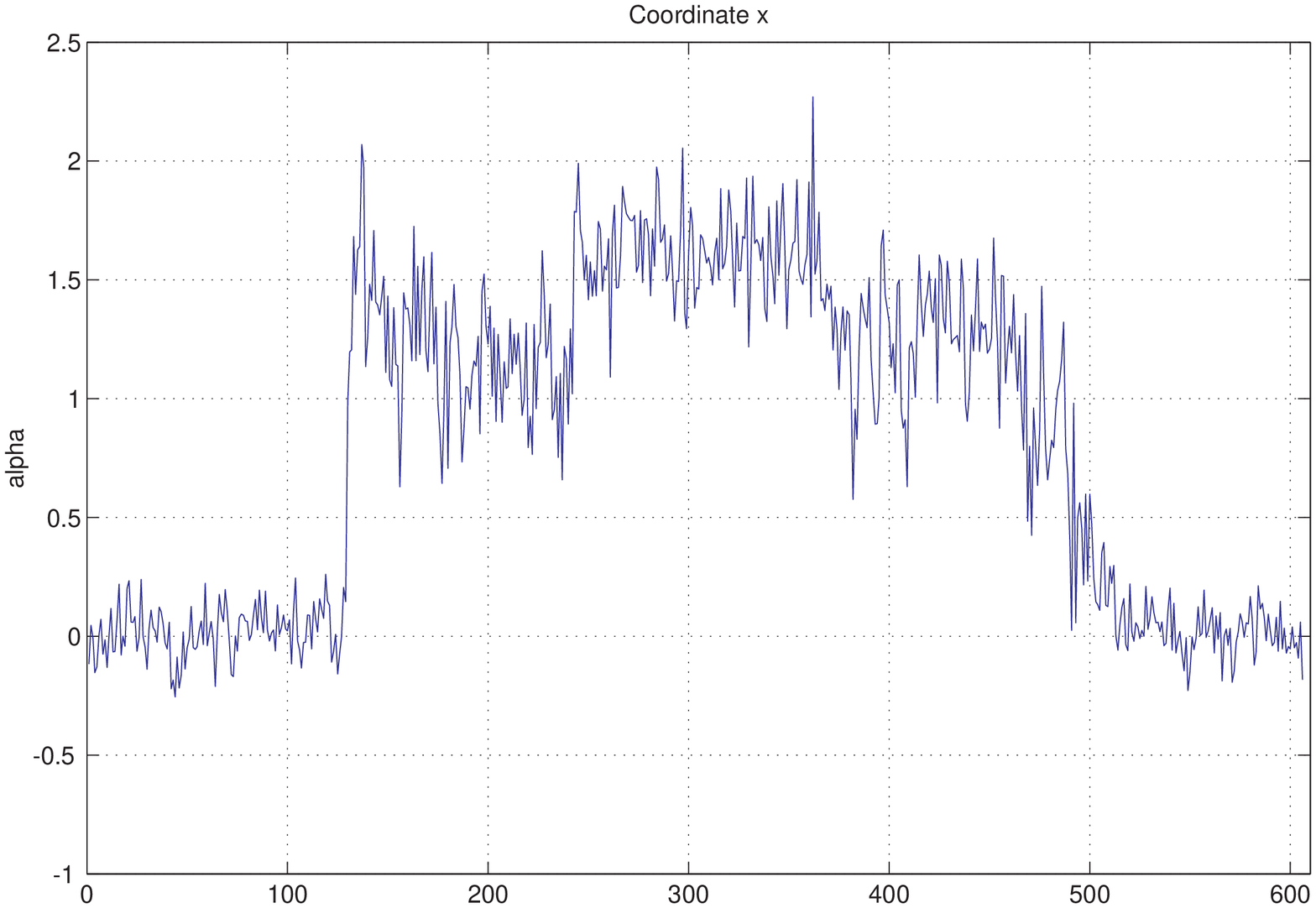}\hspace{.15cm}
          \includegraphics[height=.25\textheight]{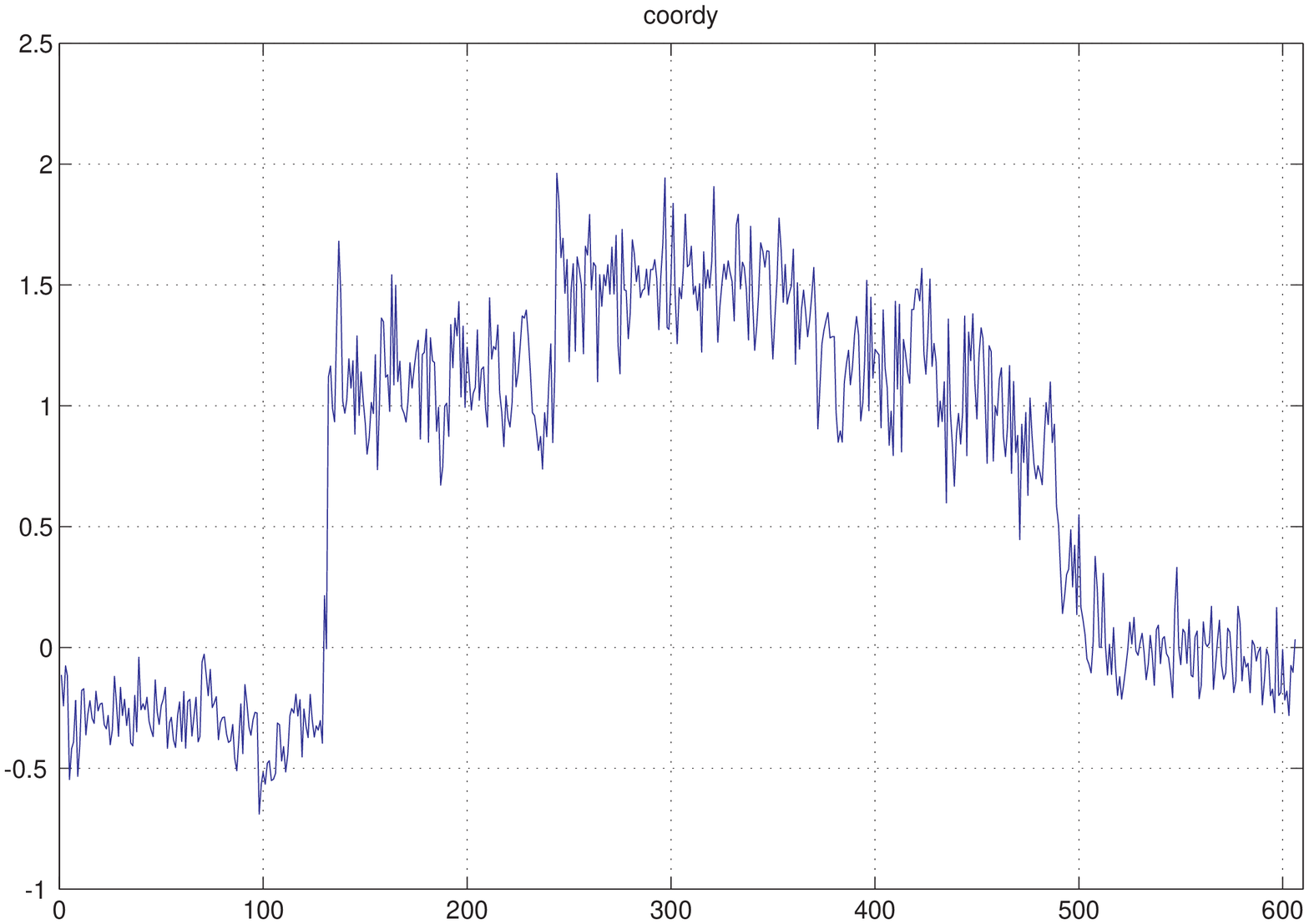}}

     \hspace{.1cm}%
     \subfigure{
          \includegraphics[height=.25\textheight]{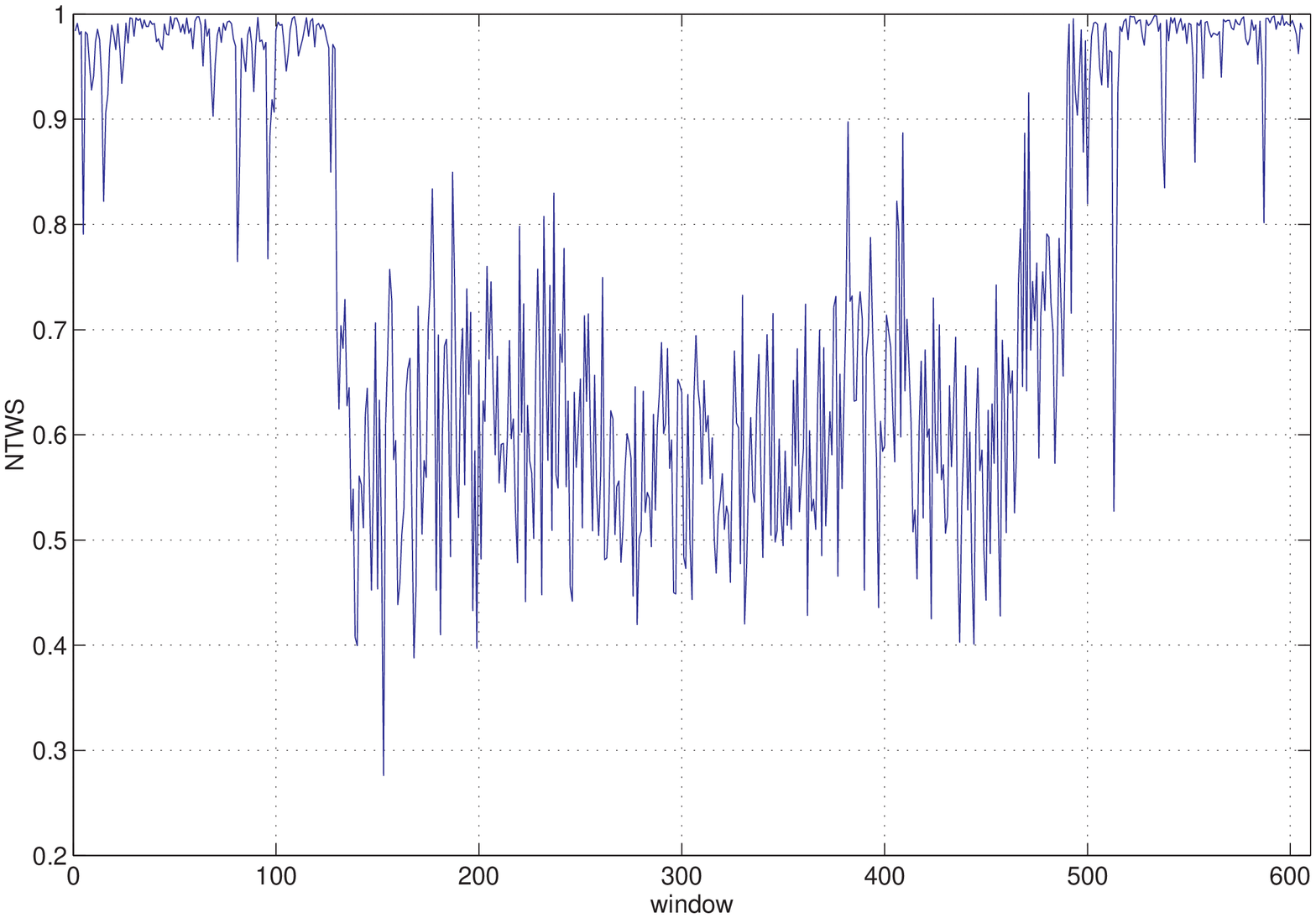}\hspace{.15cm}
          \includegraphics[height=.25\textheight]{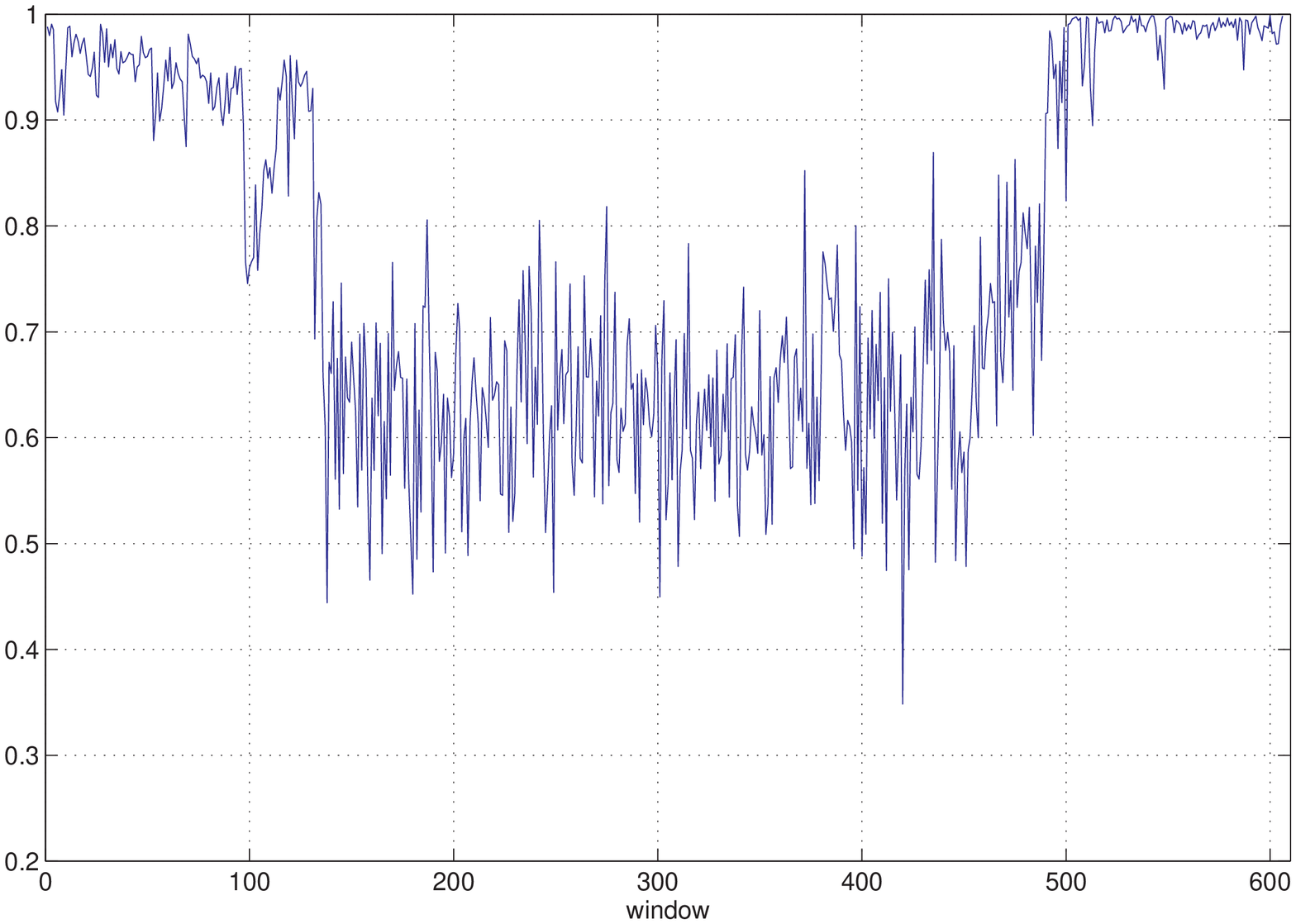}}
\caption{Quantifiers temporal evolution: $\alpha$ (top) and NTWS (bottom) for the $x$ (left) and $y$ (right) coordinates. \label{figure:3}}
\end{figure}

Figure \ref{figure:3} shows the quantifiers temporal evolution. Both quantifiers reveal that when the turbulence is normal the detector can not be able to resolve position differences, and electronic noise associated to the detector is observed. The $\text{NTWS}$ is near one as is expected for a signal generated by a totally random process and the $\alpha$ value matches with a white noise. When the turbulence is increased the system changes in an abrupt way---see coordinates' graphs at Fig.~\ref{figure:2}. It is interesting to observe the transition between the different intensities of turbulence for the signal and its corresponding quantifiers. The Hurst parameter discriminates between the other two increased turbulences. It is possible, in average, to associate a value $\alpha_x=1.17$, $\alpha_y=1.04$ for the soft case and $\alpha_x=1.62$, $\alpha_y=1.51$ for the hard turbulence. It should be noted that the signal has more regularity for the strongest turbulence. The $\text{NTWS}$ diminishes notably showing an increment in the order of the system but it is unable to distinguish between soft and hard turbulence giving values of $\text{NTWS}_x=0.63$, $\text{NTWS}_y=0.66$ for the soft turbulence and $\text{NTWS}_x=0.57$, $\text{NTWS}_y=0.63$ for the hard turbulence.

It can be followed by comparing Fig.~\ref{figure:2} and Fig.~\ref{figure:3} that the behavior of the signal is different for both coordinates. Nevertheless, the temporal evolution of the quantifiers is very similar. It is also observed that the system has hysteresis effect (see Fig.~\ref{figure:2} and Fig.~\ref{figure:3}) as it was expected. 

The mBm model is justified for modeling the dynamics associated to these processes. We conclude that the associated scaling exponent changes continuously with the turbulence strength. In the future a new generalization will be considered and studied, the \textit{generalized multifractional Brownian motion} (gmBm)~\cite{paper:ayache2004}. That processes consider that scaling exponent variations may be very erratic and not necessarily a continuous function of time. The latter condition is a strong limitation in turbulence studies where the scaling exponent can change widely from time to time.

\ack   
 
This work was partially supported by Consejo Nacional de Investigaciones Cient\'{\i}ficas y T\'ecnicas (CONICET, Argentina) and Pontificia Universidad Cat\'olica de Valpara\'iso (Project No. 123.774/2004, PUCV, Chile).

\bibliography{references}
\bibliographystyle{elsart-num}

\newpage\pagestyle{empty}

\end{document}